\documentclass[conference]{IEEEtran}
\IEEEoverridecommandlockouts

\usepackage{graphicx}
\usepackage{balance}
\usepackage{colortbl}

\usepackage{url}
\usepackage  {hyperref}

\usepackage{algorithm} 
\usepackage{algpseudocode}
\usepackage{tabularx}
\algblockdefx[Process]{Process}{EndProcess}[2][Unknown]{{\bf Process} {\it #2}}{}
\algblockdefx[Event]{Event}{EndEvent}[2][Unknown]{{\bf upon event #2 do}}{}

\usepackage{xcolor}
\usepackage{booktabs} 

\usepackage{amsmath,amssymb,amsfonts}

\makeatletter

\def\env@sqcases{%
  \let\@ifnextchar\new@ifnextchar
  \left\lbrack
  \def\arraystretch{1.2}%
  \array{@{}l@{\quad}l@{}}%
}
\makeatother

\let\proof\relax
\let\endproof\relax
\usepackage{amsthm}
 
\theoremstyle{definition}
\newtheorem{definition}{Definition}

\newtheorem{theorem}{Theorem}

\newenvironment{sketch}{%
  \proof}{\endproof}

\newcommand {\FlameStream} {FlameStream}

\begin{document}

\title {Delivery, consistency, and determinism: rethinking guarantees in distributed stream processing}

\author{\IEEEauthorblockN{1\textsuperscript{st} Artem Trofimov}
\IEEEauthorblockA{\textit{JetBrains Research} \\
Saint Petersburg, Russia \\
trofimov9artem@gmail.com}
\and
\IEEEauthorblockN{2\textsuperscript{nd} Igor E. Kuralenok}
\IEEEauthorblockA{\textit{Yandex} \\
Saint Petersburg, Russia \\
solar@yandex-team.ru}
\and
\IEEEauthorblockN{3\textsuperscript{rd}  Nikita Marshalkin}
\IEEEauthorblockA{\textit{VK} \\
Saint Petersburg, Russia \\
n.marshalkin@corp.vk.com}
\and    
\IEEEauthorblockN{4\textsuperscript{th} Boris Novikov}
\IEEEauthorblockA{\textit{National Research University}\\ \textit{Higher School of Economics} \\
Saint Petersburg, Russia \\
borisnov@acm.org}
}


\maketitle

\begin{abstract}
Consistency requirements for state-of-the-art stream processing systems are defined in terms of delivery guarantees. 
Exactly-once is the strongest one and the most desirable for end-user. 
However, there are several issues regarding this concept. 
 Commonly used techniques that enforce exactly-once produce significant performance overhead. 
Besides, the notion of exactly-once is not formally defined and does not capture all properties that provide stream processing systems supporting this guarantee. 
In this paper, we introduce a formal framework that allows us to define streaming guarantees more regularly. 
We demonstrate that the properties of delivery, consistency, and determinism are tightly connected within distributed stream processing. 
We also show that having lightweight determinism, it is possible to provide exactly-once with almost no performance overhead. Experiments show that the proposed approach can significantly outperform alternative industrial solutions.
\end{abstract}

\section {Introduction}

\label {fs-intro-seciton}

Distributed batch processing systems, such as Google's MapReduce~\cite{Dean:2008:MSD:1327452.1327492}, Apache Hadoop~\cite{hadoop2009hadoop}, and Apache Spark~\cite{Zaharia:2016:ASU:3013530.2934664}, address a need to process vast amounts of data (e.g., Internet-scale). 
The basic idea behind them is to independently process large data blocks (batches) that are collected from static datasets. 
The main advantages of these systems are fault tolerance and scalability~\cite{borthakur2011apache} of massively parallel computations on commodity hardware.

However, there are plenty of scenarios where processing results are most valuable at the time of data arrival, for example, IoT, news processing, financial analysis, anti-fraud, network monitoring, etc. 
Such problems cannot be directly addressed by classical MapReduce~\cite{Doulkeridis:2014:SLA:2628707.2628782}. 
State-of-the-art stream processing systems, such as Flink \cite{carbone2015apache}, Samza \cite{Noghabi:2017:SSS:3137765.3137770}, Storm \cite{apache:storm}, Spark Streaming~\cite{Zaharia:2012:DSE:2342763.2342773},   provide a computational model addressing these needs.

One of the most challenging tasks for streaming systems is to provide guarantees for data processing. 
Streaming systems must release output elements before processing has finished because input data is assumed to be unbounded. 

In distributed stream processing, consistency is usually described in terms of delivery guarantees: {\em at-most-once}, {\em at-least-once}, and {\em exactly-once}~\cite{carbone2015apache}. 
These guarantees describe a contract regarding {\em which data} will be processed and released in case of failures. 
Exactly-once is the strongest and the most valuable guarantee from the user perspective as it ensures that input elements are processed atomically and are not lost. These notions are seemingly simple but shadow the dependency of an output item on the {\em system state} as well as on the input item. 
Streaming systems face a need to recover computations consistently with previous input data, the current system state, and with the already delivered elements.
This requirement makes failure recovery mechanisms somewhat complicated. 

This complication is resolved by most of the existing stream processing engines. 
Flink ensures the atomicity of state updates and delivery using a protocol based on distributed transactions. 
Google MillWheel~\cite{Akidau:2013:MFS:2536222.2536229} enforces consistency between state and output elements by writing results of each operation to persistent external storage. 
Micro-batching engines like Storm Trident~\cite{apache:storm:trident} and Spark Streaming~\cite{Zaharia:2012:DSE:2342763.2342773} process data in small-sized blocks. 
Each block is atomically processed on each stage of a data flow, providing properties similar to batch processing. 
The price for exactly-once delivery is a high latency observed in these implementations (e.g., ~\cite{7530084, 7474816}).

The vast gap between the notion of exactly-once and the properties of its implementations indicates the lack of formalization. 
Misunderstandings of streaming delivery guarantees frequently cause debates and discussions~\cite{JerryPengStreamIO, PaperTrail}. Without a formal model, it is hard to observe similarities and distinctions between existing solutions and to recognize their limitations. In this work, we introduce a formal model of stream processing that captures delivery guarantees existing in most of the state-of-the-art systems.

Another property that can be easily obtained in batch processing systems but is hard to achieve in streaming engines is {\em a determinism}. 
The determinism means that repeated runs of the system on the same data produce the same results. It is commonly considered as a challenging task~\cite{Zacheilas:2017:MDS:3093742.3093921}. On the other hand, this property is desirable because it implies reproducibility and predictability. Intuitively, determinism is connected with delivery guarantees~\cite{Stonebraker:2005:RRS:1107499.1107504}, but, to the best of our knowledge, this relation has not been deeply investigated. 


In this work, we demonstrate that the property of determinism can mitigate an overhead on exactly-once. 
In order to prove the feasibility of efficient exactly-once over determinism, we introduce fault tolerance protocols on top of {\em drifting state} model~\cite{we2018adbis}. 
We show that lightweight determinism, together with the results of the formal inference provides for exactly-once for a negligible cost.

The contributions of this paper are the following: 
\begin{itemize}
    \item Formal model of a distributed stream processing  and   a   definition of  delivery guarantees 
    \item Demonstration that in non-deterministic systems providing exactly-once, the lower bound of latency is the duration of state snapshotting
    \item Techniques for lightweight implementation of exactly-once guarantee on top of a deterministic engine
    \item Study of the practical feasibility of the proposed approaches
\end{itemize}

The rest of the paper is structured as follows: 
in section~\ref{fs-preliminaries} the notion of consistency applied to a stream processing is discussed, 
we introduce our formal framework in section~\ref{fs-formalism}, 
existing implementations of exactly-once in terms of the proposed formal framework are described in section~\ref{fs-eo-impl}, 
implementation details of exactly-once over determinism are mentioned in section~\ref{fs-consistency-section}, 
experiments that demonstrate the feasibility of the proposed concept are detailed in section~\ref{fs-experiments-seciton}, 
and we discuss prior works on the topic in section~\ref{fs-related-seciton}. 

\section {Motivation}
\label{fs-preliminaries}

The concept of {\em consistency} is traditionally expressed in terms of the transactional behavior, known as the ACID properties of transactions in the databases. 
 These properties ensure that the (database) state is consistent to the degree required by the specified isolation level, providing consistency of processing. 
 However, database systems are not used standalone: they interact with client applications.
  From the client perspective,   the notion of consistency also includes {\em delivery guarantees} as well. {\em At-least-once} ensures that input data are not lost, {\em at-most-once} eliminates duplicate processing, and {\em exactly-once} combines both ensuring the absence of input data losses and repeated delivery of results. An implementation of exactly-once processing of transactions based on persistent queues is described in~\cite{DBLP:books/mk/WeikumV2002}.



Delivery guarantees are typically not considered in batch processing systems, because they always ensure atomicity between reading input data, processing, and delivery of results. In other words, it means that each record within a batch is processed exactly-once. This is achieved by consecutive processing of batches, persistent storage for all intermediate results, and reprocessing of all suspected failures. 
A disadvantage of this approach is the impossibility to deliver any results while the processing is not complete.

In contrast,   consistency guarantees for stream processing systems are commonly declared in terms of delivery only. 
A pitfall here is that terms like {\em exactly-once} and {\em at-least-once} hide the fact that system state must also be consistent with input and output in order to achieve correct results. 
In the absence of formal definitions, such terms can lead to an improper perception that the state does not play an essential role in consistency enforcement. 

The problem shows up if some streaming elements do not commutate within an operation in a data flow. 
Let us consider a data flow with an operation that concatenates input strings and delivers it after each item. 
The system must restore its state  (a concatenation of strings in this example) after a failure. 
A straightforward approach to restoring the state is to replay missing input elements. 
However, these elements can be reordered in an asynchronous distributed environment, potentially affecting the concatenation of input items processed exactly-once but inconsistent with output released before the failure. 
For example, if input elements are strings $"a","b","c"$, and user have already received  $"a"$ and $"ab"$, it is expected that the current state is $"ab"$, and the next output element is $"abc"$. However, after recovery and state reprocessing, the current state can become $"ba"$ due to races. In this case, the next output element $"bac"$ will be inconsistent with the previously delivered elements $"a"$ and $"ab"$.  

In real-life applications, concatenation can be faced in user behavior analysis, where the most recent actions are stored. 
This simple example illustrates that straightforward definitions of delivery guarantees are not sufficient to provide output consistency. 
While the state snapshotting protocol for distributed systems~\cite{Chandy:1985:DSD:214451.214456} was adopted for streaming systems~\cite{2015arXiv150608603C}, to the best of our knowledge, there are no formal definitions of delivery guarantees and consistent results. 
In the next section, we introduce a formal framework that allows us to define delivery guarantees with aware of the system state, data producers, and data consumers. 
We also reveal an approach to exactly-once implementation with low performance overhead.

=\section {Formal Model}
\label{fs-formalism}

In this section, we introduce a formal model of a distributed stream processing system. After that, we define the notions of delivery guarantees in a regular way. We demonstrate that in a general case, non-deterministic systems must save results of non-commutative operations before output delivery. It is illustrated how state-of-the-art stream processing systems fit the requirements that we formulate.

\subsection{Preliminaries}

A distributed stream processing system is a shared-nothing distributed runtime, that can handle a potentially infinite sequence of input items. Each item can be transformed several times before the final result is released from the system. Output elements may depend on multiple input ones. Elements can be processed one-by-one or grouped into small input sets, usually called {\em micro-batches}. 

A user specifies required stream processing with a {\em logical graph}. Vertices of this graph represent operations and edges determine the data flow between tasks. A processing system maps the logical graph to a {\em physical} graph that is used to control actual distributed execution. Commonly, each logical operation is mapped to several physical tasks that are deployed to a cluster of computational units connected through a network. Each operation may be {\em stateless} or {\em stateful}. A system is usually responsible for state management in order to prevent inconsistencies.

An input element has {\em entered} if the system is aware of this element since that moment and takes some responsibility for its processing. 
This concept can be implemented differently. 
For example,
 an  element has been entered when  it  has arrived at {\em Source} vertex in Flink, while   
an element enters, when it is read or received by an input agent also called  {\em Source}   in Spark Streaming.

An output element has {\em left} the system if the element has been released to a consumer. 
Since that time, the system cannot observe it anymore. This concept can also be implemented differently in various systems. For instance, in Spark Streaming element leaves when it is pushed to output operation, e.g., written to HDFS or file. In Flink, an element is delivered to end-user when it leaves from {\em Sink} vertex.   

It is important to note that input and output elements cannot be directly matched due to the possibility of complex transformations within the system. 
For instance, a single input element can be transformed into multiple ones.  The resulting elements may be processed in entirely different ways and even influence each other. 
In general, it is hard to find out input elements on which an output element depends. 

\subsection{Distributed streaming model}

A formalization of streaming consistency guarantees we begin with a regular definition of a stream processing system. In order to be sufficient for the formalization, such a model should cover user-defined transformations, system state, data producers, and data consumers.

Index $\tau\in{\mathbb{N}}$ can be considered as an exact global discrete time. We assume that only one event can happen at any single point in time $\tau$.

\begin{definition}{Stream processing system}
\label{reference_system}
is a tuple of $(\Gamma,D,F)$, where $\Gamma$ is a set of all possible data flow elements, $D\subseteq{2^{\Gamma}\times2^{\Gamma}}$ is a binary relation on its power set. $F$ is a recovery function that restores the working set in case of system failures. Computations within the system are defined by the recurrent rules on the set of input elements $A$, output elements $B$, and the transient working set $W$. On each iteration, one of the following steps is chosen:

\begin{enumerate}
    \item \textbf{Input} $a_\tau\in{\Gamma}$:\\ $A_{\tau+1}=A_\tau \cup \{a_\tau\}, \\ B_{\tau+1}=B_{\tau}, \\ W_{\tau+1}=W_{\tau} \cup \{a_\tau\}.$
    \item \textbf{Output} $b_\tau\in{W_\tau}$:\\ $A_{\tau + 1} = A_{\tau}$, \\ $B_{\tau+1}=B_\tau \cup \{b_\tau\}, \\ W_{\tau+1}=W_{\tau} \setminus \{b_\tau\}.$
    \item \textbf{Transform}\\ $A_{\tau + 1} = A_{\tau}$,\\ $B_{\tau+1}=B_{\tau}$, \\ $W_{\tau+1}=(W_\tau \setminus \chi_\tau) \cup Y_\tau, (\chi_\tau,Y_\tau) \in D$, \\where\\$\chi_\tau \thicksim P(X\mid X \subseteq W_\tau \cup B_\tau , \exists Y : (X,Y) \in D)$, \\ probability distribution of $\chi_\tau$ depends on a system architecture. \label{random_formula}
    \item \textbf{Failure and recover} \\  $A_{\tau + 1} = A_{\tau}$,\\ $B_{\tau+1}=B_{\tau}$, \\ $W_{\tau+1} = F(A_\tau,B_\tau)$
\end{enumerate}

\end{definition}

\begin{table}[!b]
    \caption{Notations used throughout the paper}
    \begin{tabular}{l|p{5cm}}
        \hline
        $\Gamma$ & The set of all possible data flow elements \\ 
        \hline
        $D\subseteq{2^{\Gamma}\times2^{\Gamma}}$ & Binary relation that captures user-defined operations  \\
        \hline
        $Cl_D$ & Transitive closure of $D$  \\
        \hline
        $Cl^{-1}_D$ & Inversed transitive closure of $D$  \\
        \hline
        $\tau \in \mathbb{N}$ & Exact global discrete time \\
        \hline
        $a_\tau \in \Gamma$ & Input element at the time $\tau$ \\
        \hline
        $A_\tau \subseteq \Gamma$ & All input elements by the time $\tau$ \\
        \hline
        $b_\tau \in 2^{\Gamma}$ & Output element at the time $\tau$ \\
        \hline
        $B_\tau \subseteq 2^{\Gamma}$ & All output elements by the time $\tau$ \\
        \hline
        $W_\tau \subseteq 2^{\Gamma}$ & Working set at the time $\tau$ \\
        \hline
        $F$ & Recovery function \\
        \hline
        $F^{*}$ & Reference recovery function \\
        \hline
        $P(b_{\tau+1}|A_{\tau}, B_\tau, F)$ & Probability of output element \\
        \hline
        $P(b_{\tau+1}|A_{\tau}, B_\tau, F^{*})$ & Probability of output element in a system with a reference recovery function \\
    \end{tabular}
    \label{notations}
\end{table}

Data producers and consumers are modeled using sets $A$ and $B$, respectively. All elements that in the system at the time $\tau$ are also presented in the set $W_\tau$. We assume that operation states are common elements that are presented in $W_\tau$ as well. Binary relation $D$ captures all possible transformations within a physical graph. {\em Input} step indicates that a new streaming element enters the system. This element is also presented in set $A$, because the system may request it for reprocessing. {\em Output step} describes a case, when an element leaves the system, e.g., during delivery of an output element to a data consumer. {\em Transform} step denotes the internal transformation of an element according to user-defined operations. In this model, we do not explicitly introduce computational units and asynchronous network channels. Instead, we simulate possible races and distributed asynchronous processing through the probability of the next transformation inside a data flow. Such probability is modeled using a random variable $\chi_\tau$, which indicates the input of an operation that will be executed next. {\em Failure and recover} step indicates the recovery of the system state based on input and output elements.


We allow using output elements in system transformations because we suppose that {\em snapshots} of operation states taken by real stream processing systems are common output elements. The rationale for this assumption is the following:

\begin{itemize}
    \item Snapshots should be considered as an output as they are usually stored in external systems, (such as in HDFS, Kafka, relational databases)   and are read back during recovery. 
    \item Stream processing systems aim at minimizing information which is needed to recover processing, but their optimizations do not directly affect consistency. The notion of $F(A_\tau, B_\tau)$ makes our concept general enough to describe any streaming engine while being independent of a specific implementation of $F$.
\end{itemize}

\begin{figure}[htbp]
  \centering
  \includegraphics[width=0.48\textwidth]{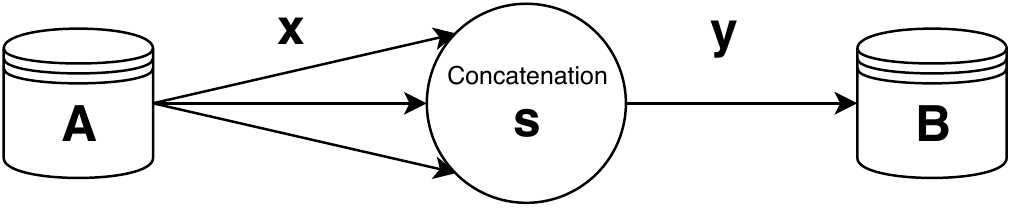}
  \caption{Strings concatenation physical graph}
  \label{concat}
\end{figure}

Let us illustrate the proposed model by an example. Assume that a physical graph consists of a single windowed operation (window = 2) that concatenates strings from multiple asynchronous input channels. This graph is illustrated in Figure~\ref{concat}: $x \in W$ indicates input channels, $y\in W$ denotes output channel, and $ s \in W$ is the state of the operation. Set $\Gamma$ contains all possible characters, and binary relation $D$ contains all possible options for their concatenation. Recovery function $F$ reprocess all input elements since the last snapshot. Table~\ref{concat_example} demonstrates a potential execution of the defined graph with strings $"a","c","b","d","e"$ as input elements. The concatenation operation receives a random element from its input due to asynchronous channels. Wall time is indicated to highlight that races may occur in a data flow. Note that step {\em snapshot} can be considered as an ordinary output step. Red lines emphasize the recovery process after failure. 

As one can see, the system reaches precisely the same state after recovery. Eventually, the end-user receives the output element {\em be} that is expected within such execution. Further, we demonstrate that simple reprocessing of missed input elements can cause significant inconsistencies in results.



\begin{table}[htbp]
\begin {center}
\caption{String concatenation data flow}   \tabcolsep2pt
\begin{tabular}{c|l|c|c|c|c|c|c}      
wall\\ time & $\tau$& $a\in A$  &$x\in W$& $s\in W$ & $y\in W$ & $b \in B$ & step  \\
\hline
1   &   1   &   a   &   a           &           &       &           &   input \\
1   &   2   &   c   &   a,c         &           &       &           &   input \\
1   &   3   &       &    c          &   [a]     &    a  &           &   transform \\
1   &   4   &       &   c           &   [a]     &       &   a       &   output \\
1   &   5   &       &               &   [a,c]   &   ac  &           &   transform \\
1   &   6   &       &               &   [a,c]   &       &   ac       &   output \\
1   &   7   &       &               &   [a,c]   &       &   [a,c]   &   snapshot \\
2   &   8   &   b    &   b          &   [a,c]   &       &           &   input \\
2   &   9   &   d    &   b,d        &   [a,c]   &       &           &   input \\
2   &   10   &       &   b          &   [c,d]   &  cd   &           &   transform \\
2   &   11   &       &   b          &   [c,d]   &       &      cd     &   output \\
2   &   12   &       &              &   [d,b]    &  db   &           &   transform \\
2   &   13   &       &              &   {\bf [d,b]}    &       &      {\bf db}     &   output \\
\arrayrulecolor{red}\hline
3   &   14   &       &               &   [a,c]         &       &           &   recovery \\
3   &   14   &   b   &   b           &   [a,c]         &       &           &   \\
3   &   14   &   d   &   b,d         &   [a,c]         &       &           &   \\
3   &   14   &       &   b           &   [c,d]         &       &           &   \\
3   &   14   &       &   b           &   [c,d]         &       &           &   \\
3   &   14   &       &               &   [d,b]         &       &           &    \\
3   &   14   &       &               &   {\bf [d,b]}   &       &           &    \\
\arrayrulecolor{red}\hline
4   &   15   &  e    &   e           &   [d,b]   &      &           &   input \\
4   &   16   &       &              &   [b,e]   &   be    &          &   transform \\
4   &   17   &       &              &   [b,e]   &       &   {\bf be}       &   output \\
\end{tabular}
\label{concat_example}
\end {center}
\end{table}

\subsection{Model analysis}

In a classical model proposed by Chandy and Lamport~\cite{Chandy:1985:DSD:214451.214456}, a distributed system is represented as a graph of processes, which can be connected via channels. Each process can own a modifiable internal state, generate {\em events} and send them to other processes through the channels. {\em Global system state} in this model contains processes states and channel states, e.g., elements, which are in-flight at the moment. Distributed asynchronous processing is simulated using permutations of events.

The proposed model is a variation of a classical Chandy-Lamport distributed system model with the following properties and modifications:

\begin{itemize}
    \item Our notion of a streaming system does not provide a concept of {\em operations state} because, as it is shown in~\cite{we2018adbis}, states can be considered as common data items. In this case, a state element is presented in a system until it is transformed into a new state together with new elements. This model allows a binary relation $D$ to capture computations within a graph of processes that is a physical streaming execution graph in our case. Hence, the working set $W_\tau$ is a global state with only channel states in terms of Chandy-Lamport.
    \item Input and output elements $a_\tau$ and $b_\tau$ are special events. This way, {\em storage} and {\em computations} are distinguished. We assume that output elements $B_\tau$ are all data that leaves a system including so-called {\em state snapshots} which are typically used for recovery~\cite{Carbone:2017:SMA:3137765.3137777}. Because of this, we state that $X \subseteq W_\tau \cup B_\tau$ in the transform transition of the working set. In most state-of-the-art stream processing systems, the end-user is external to the system, i.e., a user can observe input and output elements, but not the working set.
    \item Distributed asynchronous computations are modeled through the random choice of an operation that is executed next. In the proposed model, instead of using permutations of data flow elements that fixate an execution route, we suppose that the next transition of the working set is a random variable. The distribution of this variable depends on system architecture. 
\end{itemize}

\subsection{Consistency}

The parallel asynchronous nature of distributed stream processing allows us to handle only a {\em probability} of an output element because it can be different from run to run due to element reorderings.

\begin{definition}{Probability of output element}
$P(b_{\tau+1}|A_{\tau}, B_\tau, F)$ is a probability to observe output element $b_{\tau+1}$ considering all previous input and output elements, and recovery function.
\end{definition}

The problem here is that it is hard to define the correct output, because it may significantly vary. Inconsistencies may arise only in case of failure and incorrect recovery. Let us introduce a {\em reference} recovery function that allows us to denote results which are possible to reach without failures.

\begin{definition}{Reference}
$F^{*}$ is a recovery function that restores exactly the same working set as before the failure, $\forall \tau \in \mathbb{N}, F^{*}(A_\tau,B_\tau)=W_\tau$.
\end{definition}

\begin{definition}{Probability of output element with a reference recovery function}
$P(b_{\tau+1}|A_{\tau}, B_\tau, F^{*})$ is a probability to observe output element $b_{\tau+1}$ considering all previous input and output elements, and a reference recovery function.
\end{definition}

Using the notion of a reference recovery function, one can define the correct output. 

\begin{definition}{Output element $b_{\tau+1}$ is consistent}
with all previous input and output elements $A_\tau$ and $B_\tau$ if the probability to observe it is non-zero in a system with reference recovery: $P(b_{\tau+1} \mid A_\tau,B_\tau,F^{*})>0$.
\end{definition}

The reference recovery function allows us to express the notion of correct execution in terms of correspondence between input and output elements. In most real cases, input and output elements are the only data that can be observed by the end-user. In real distributed stream processing systems, failures and recoveries can corrupt the output, despite the fact, that in terms of a simple definition of delivery guarantees, all elements are processed exactly-once. 

The problem here is that elements can be concatenated in a different order on recovery. Let us demonstrate it by the already mentioned example demonstrated in Figure~\ref{concat} and Table~\ref{concat_example}. During recovery system may reach concatenation state $s=[b,d]$ at $\tau=15$ rather than $s=[d,b]$ due to asynchronous input. In this case, output at $\tau=17$ becomes {\em de} instead of {\em be}. However, {\em de} is inconsistent with the previous output, because element with prefix {\em d} has already been released at $\tau=13$. It is important to note that all input elements within the example are applied to concatenation state {\em  exactly-once}. This simple example demonstrates that informal definition of exactly-once does not guarantee the consistency of output elements.

\subsection{Delivery guarantees}

As we demonstrate further, state-of-the-art stream processing systems prevent inconsistencies illustrated in the example above. However, the informal definition of exactly-once is insufficient to describe the level of data consistency that is provided. Here we introduce regular definitions of delivery guarantees which take this issue into account. 

\begin{definition}{System provides for exactly-once}
if it is possible to obtain each output element $b_{\tau+1}$ in a system with a reference recovery fucntion, i.e.,\\ 
$\forall{\tau \in \mathbb{N}}, b_{\tau+1}\in 2^{\Gamma}: P(b_{\tau+1}|A_{\tau},B_\tau,F)>0 \Rightarrow \\ P(b_{\tau+1}|A_{\tau},B_\tau,F^{*})>0$.
\end{definition}

\begin{definition}{System provides for at-most-once}
if \\
$\exists{A^{0}_{\tau}\subseteq{A_{\tau}}}$ such that \\
$\forall{\tau \in \mathbb{N},{b_{\tau+1}\in 2^{\Gamma}}}: P(b_{\tau+1}|A_{\tau},B_\tau,F)>0 \Rightarrow \\ P(b_{\tau+1}|A^{0}_{\tau},B_\tau,F^{*})>0$.
\end{definition}

\begin{definition}{System provides for at-least-once}
if \\
$\exists{A^{*}_{\tau}\subseteq{2^{A_{\tau}}}}$ such that \\
$\forall{\tau \in \mathbb{N}, {b_{\tau+1} \in 2^{\Gamma}}}: P(b_{\tau+1}|A_{\tau},B_\tau,F)>0 \Rightarrow \\ P(b_{\tau+1}|A^{*}_{\tau},B_\tau,F^{*})>0$.
\end{definition}

Exactly-once states that observed results cannot be distinguished from one of the possible results produced by a system with a reference recovery function. It means that even if data flow contains races and some elements do not commute, output consistency will be preserved. Looking back to the example illustrated in Table~\ref{concat_example}, our formal definition of exactly-once ensures that after recovery system reaches state $[d,b]$, not $[b,d]$, because this state has already influenced output elements.

At-most-once and at-least-once guarantees are the relaxations of exactly-once. The results within these guarantees can be obtained without failures, but with the modification of the input. At-least-once can be reproduced if the input contains duplicated items. At-most-once can be achieved with a reference recovery if some input elements are missed. It is important to note, that regarding at-most-once guarantee we require an input element to be processed atomically with all its derivatives or not processed at all. To the best of our knowledge, none of the stream processing engines supports at-most-once guarantee, so we cannot verify the relevancy of this assumption. It is accessible to provide at-most-once by producing no output at all, but if the output is presented, at-most-once enforcement becomes much more difficult.  

In our formal model, the relaxations of exactly-once are defined without diving into recovery mechanisms. Instead, they are described through possible input channel flaws in a system with a reference recovery. This trick allows us to represent invisible system details in terms clear for an external user.

\subsection{Exactly-once and determinism}

Operations that are presented in $D$ can directly affect the complexity of consistency enforcement mechanisms. For example, if all operations are idempotent,  exactly-once can be achieved~\cite{Akidau:2013:MFS:2536222.2536229}. Among common types of operations, non-commutative ones provide significant difficulties in achieving exactly-once.

\begin{definition}{D contains non-commutative operation}
if\\ 
$\exists (x,y), s_1, s_2 \in \Gamma, s_1 \neq s_2: \\ ((x,y),s_1)\in D, \\ ((y,x),s_1)\notin D, \\ ((y,x),s_2)\in D$.
\end{definition}

The are many non-commutative operations that are commonly used in practice: concatenation, matrix multiplication, cross product, etc. Hence, general-purpose stream processing systems should support them. The problem with such operations we demonstrated above: simple reprocessing of missing elements in case of failure may lead to results inconsistencies.

\begin{definition}{System is deterministic}
if\\ 
$\forall{\tau\in{\mathbb{N}}, b_{\tau+1}\in{2^{\Gamma}}}:P(b_{\tau+1}|A_{\tau},B_\tau,F^{*})=1$.
\end{definition}

An essential property of a deterministic system is that it preserves the same order of elements before non-commutative operations on each run. We further demonstrate how this property can be used to achieve exactly-once with low performance overhead.

On the other hand, the absence of determinism imposes restrictions on exactly-once implementations. The following theorem denotes the necessary and sufficient conditions for exactly-once in non-deterministic systems with a non-commutative operation in $D$. It demonstrates that if a system is non-deterministic, but supports non-commutative operations, it must save (take a snapshot of) results of non-commutative operations before delivery of output elements that depend on these results.

\begin{figure}[htbp]
  \centering
  \includegraphics[width=0.48\textwidth]{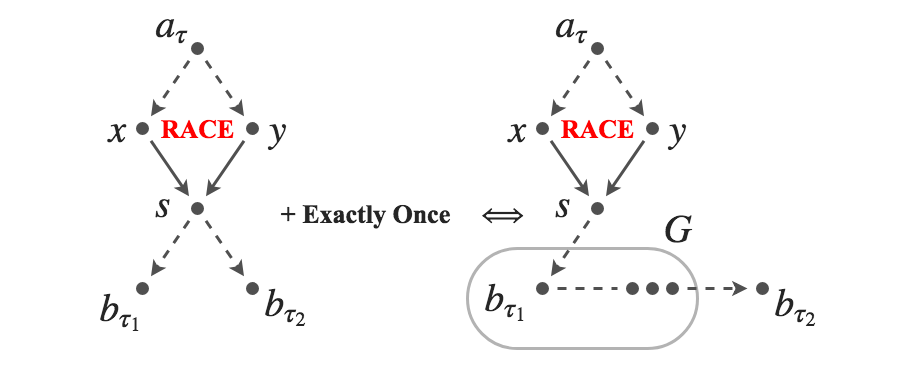}
  \caption{The scheme of Theorem~\ref{necessary_conditions}}
  \label {theorem-pic}
\end{figure}

\begin{theorem}
\label{necessary_conditions}
If $D$ contains non-commutative transformation and has the following dependencies for some $a_\tau \in \Gamma$:

\begin{enumerate}
  \item[(i)] $x \in Cl_D(a_\tau), y \in Cl_D(a_\tau)$
  \item[(ii)] $((x,y),s)\in D$ through a non-commutative operation
  \item[(iii)] $b_{\tau_1}, b_{\tau_2} \in Cl_D(s), \tau_1 < \tau_2$
\end{enumerate}

\noindent then non-deterministic system supports exactly-once if and only if:

\begin{equation}
\label{theorem_conditions}
  \begin{gathered}
    G = Cl_D(b_{\tau_1}) \cap C_D^{-1}(b_{\tau_2}) \\
    \forall (u, v) \in D, u \in Cl_D(s): v \subset G \Rightarrow u \subset G
  \end{gathered}
\end{equation}

\end{theorem}
\begin{sketch}
$ $\newline

The scheme of the theorem is shown in Figure~\ref{theorem-pic}. The theorem conditions mean that each result of the non-commutative operation must become recoverable before the delivery of output that depends on this result. 
$ $\newline

\noindent{\em Necessary condition}\newline

  Assume that conditions~\ref{theorem_conditions} are not satisfied, i.e. $\exists v \subset G, u \not\subset G: (u, v) \in D$, and system fails at time $\tau_1<\tau_f<\tau_2$. Then, to obtain $b_{\tau_2}$ there is a need to recompute $v$, $u$ and therefore $s$. However, due to asynchronous distributed processing and non-commutative transformation, system can reach not exactly $s$ such that $((x,y), s) \in D$, but $s':((y,x),s')\in D$. In this case, after recovery element $b_{\tau_2}$ will be delivered, but it will depend not on $s$, but on $s'$: $b_{\tau_2}\in Cl_D(s')$. This is a contradiction, because $b_\tau$ has been already delivered and $P(b_{\tau_2}\in Cl_D(s')|\{a_\tau\},\{b_{\tau_1} \in Cl_D(s) \})=0$.
$ $\newline

\noindent{\em Sufficient condition}
\newline

Let us assume that a  system fails at time $\tau_f$. If $\tau_f < \tau_1$ or $\tau_f > \tau_2$,  exactly-once is obviously satisfied. Assume that $\tau_1<\tau_f<\tau_2$. In this case, $b_{\tau_1}\in B_{\tau_f}\subset B_{\tau_2}$. Hence, $b_{\tau_2}$ can be restored directly from $b_{\tau_1}$ without reprocessing of s, i.e. $F(a_\tau,b_{\tau_1})=b_{\tau_2}$ and $b_{\tau_1}, b_{\tau_2} \in Cl_D(s)$ after recovery, thus $P(b_{\tau_2}\in Cl_D(s)|\{a_\tau\},\{b_{\tau_1} \in Cl_D(s) \})>0$.
\end{sketch}

We assume   in this  theorem    that input element $a_\tau$ is split into elements $x,y$. The same behavior can be reproduced if two input elements $a_\tau$ and $a_{\tau+1}$ enter a system through asynchronous channels. In general, this behavior is natural for stream processing systems because elements are processed one-by-one without synchronization and order enforcement. 

This theorem has direct practical implications for non-deterministic systems:
\begin{itemize}
    \item {\bf Output elements must wait until the snapshot is taken}. If a system aims at providing exactly-once, it must output elements only if there exists a snapshot that contains results of non-commutative operations. The problem here is that a system is usually not aware of user-defined operations and cannot distinguish commutative and non-commutative operations, so it must wait even if all elements commutate.
    \item {\bf Latency is affected by the period of snapshotting}. The consideration above implies that the lower bound of latency in the worst case in non-deterministic systems is the snapshotting period, together with the duration of taking a snapshot. There is a trade-off between latency and the frequency of taking snapshots because too frequent snapshotting can cause high extra load, while rare snapshots lead to high latency.
\end{itemize}

It is important to note that the proposed model is suitable not only for the formal analysis of the properties of exactly-once but for a deeper understanding of the other aspects of stream processing. While these topics are promising as well, they are out of the scope of this paper.

\section {Exactly-once implementations}
\label{fs-eo-impl}


\subsection{MillWheel}

 The $\Gamma$ is all possible streaming {\em records}. Dependency relation $D$ is defined using a graph of user-defined transformations. 
 To ensure fault tolerance,   MillWheel employs a {\em strong productions}~\cite{Akidau:2013:MFS:2536222.2536229}   mechanism that stores persistently input and output of operation for each new input item.
 Thus each streaming element is saved before it may influence other elements or operation states. 
 Recovery function $F$ resends all saved records in case of failure, and each operation deduplicates input that has been already processed.

\begin{figure}[htbp]
  \centering
  \includegraphics[width=0.48\textwidth]{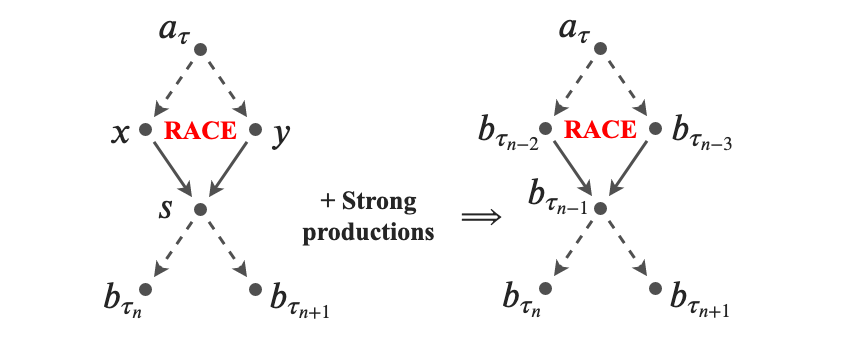}
  \caption{Strong productions mechanism}
  \label{millwheel}
\end{figure}

The necessary condition of exactly-once from Theorem~\ref{necessary_conditions} claims that each element $s \in \Gamma$ obtained through a non-commutative operation must be persistently saved before any other element $b_{\tau} \in Cl_D(s)$ that depends on $s$ is released. Strong productions mechanism ensures that this condition is satisfied, i.e., if some element $b_\tau$ is released, it is guaranteed that its dependencies were released earlier (saved to persistent storage). The idea of this method is shown in Figure~\ref{millwheel}. It demonstrates that in MillWheel, all elements become recoverable, so there is no need to reprocess them again from an input item in case of failure. 

Such behavior is also called {\em effective determinism}~\cite{akidau2018streaming} because computations become deterministic until the persistent storage is cleared. Strong productions method guarantees that even if many possible results of a non-commutative operation may be obtained due to races, only one of them is computed and saved. The price for such exactly-once enforcement is an overhead on external writes on each transformation in a physical graph.

\subsection{Spark streaming}

In Spark, the  $\Gamma$   is a set of all possible small collections of input items called  {\em RDD} records that are processed atomically.
Dependency relation on the power set of $\Gamma$ is also defined using a directed acyclic graph of user operations. 
Spark inherits main properties of batch processing systems:  each new stage of computations is started only after the previous one is completed and recovery function $F$ starts reprocessing of a micro-batch from the beginning in case of failure.

\begin{figure}[htbp]
  \centering
  \includegraphics[width=0.48\textwidth]{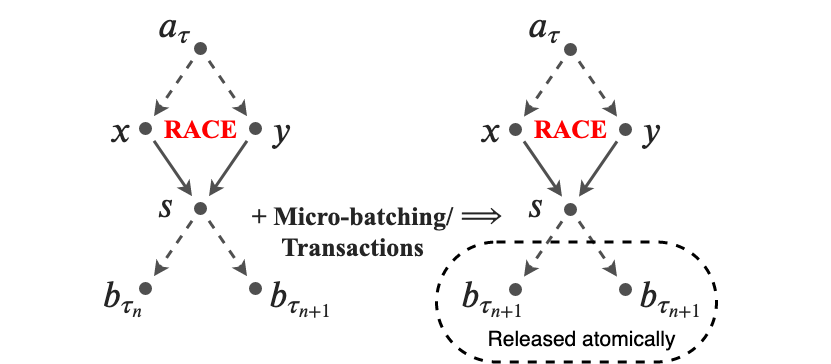}
  \caption{Micro-batching and transactional approach}
  \label{spark_flink}
\end{figure}

The condition for exactly-once from Theorem~\ref{necessary_conditions} is satisfied because all elements within the micro-batch are released atomically. Elements from various micro-batches can interact only through persistently stored state. 
Output elements that are ready to be delivered to the end-user must wait until the whole micro-batch is completely processed. 
Therefore, if two output elements $b_{\tau_1},b_{\tau_2} \in Cl_D(s)$ depend on a single item $s$, it is guaranteed that $\tau_1=\tau_2$. The illustration of this behavior is shown in Figure~\ref{spark_flink}. 
 
\begin{figure*}[tbp]
  \centering
  \includegraphics[width=0.78\textwidth]{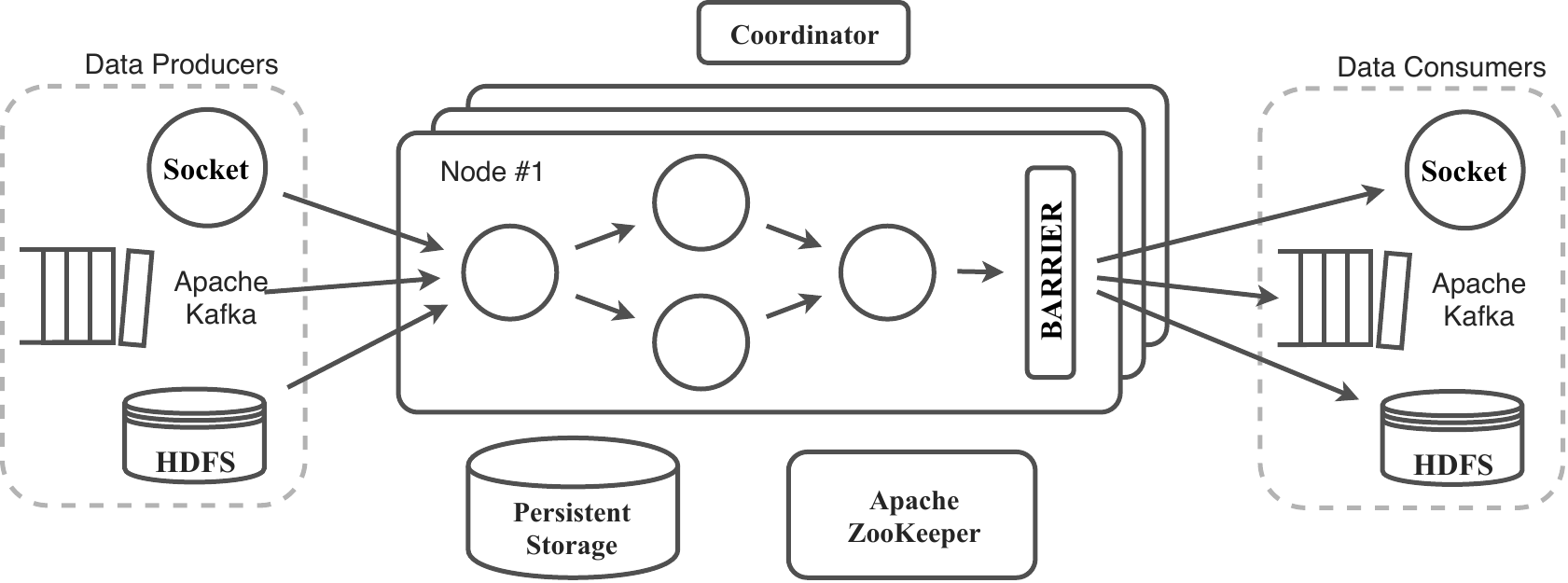}
  \caption{The overview of an architecture for drifting state implementation}
  \label {arch}
\end{figure*}
 

\subsection{Flink}

In Flink, the  $\Gamma$ is represented as all {\em StreamRecords}. Relation $D$ is defined in the form of a directed acyclic graph consisting of user operations. 
Flink periodically saves information needed for recovery by injecting particular elements called {\em checkpoints} into the input stream. 
Intervals  between checkpoints are called {\em epochs}. Checkpoints go through the same network channels as common elements and push all inverted dependencies of inputs through the system. 
Each operation prepares data to save independently at the moment of checkpoint arrival. Prepared data is committed to external storage when checkpoint passes through the whole data flow. Output elements are delivered to the end-user only after the commit. 
Recovery function $F$ reprocesses last non-complete epochs.

This mechanism ensures that all elements within an epoch are processed atomically and all interactions between elements from different epochs are possible only through persistent state. 
Hence, as it is shown in Figure~\ref{spark_flink},
 if two output elements $b_{\tau_1},b_{\tau_2} \in Cl_D(s)$ depend on a single item $s$, 
 it is guaranteed that $\tau_1=\tau_2$. 
This ensures that the necessary condition of exactly-once from Theorem~\ref{necessary_conditions} is satisfied. 
Such a method is quite similar to the micro-batching technique used in Spark. 
The differences are: elements are buffered after their processing is done and  multiple epochs can be processed simultaneously. 


\section{Exactly-once over determinism}

\label {fs-consistency-section}

In the previous section, we demonstrated how state-of-the-art stream processing systems handle non-deterministic behavior in order to achieve~ exactly-once. All these systems make state recoverable before output elements which depend on this state are released. As we demonstrate further, such behavior can significantly increase the processing latency of individual elements.

The condition from the Theorem~\ref{necessary_conditions} can be relaxed for a deterministic system because determinism guarantees order enforcement before non-commutative operations. If a system is deterministic, it is possible to recompute states of non-commutative operations consistently. Hence, such a system can release an output element before a snapshot is taken because it can reproduce the same state in case of failure. It means that even if elements are divided into the epochs like in Apache Flink, and a snapshot is taken once per epoch, ready-to-release output elements should not wait until the epoch is committed in order to be delivered. The limitation of this approach is that only deterministic operations (e.g. without the usage of random values) are allowed in a data flow.

To combine the determinism with  exactly-once we need to design protocols for recovery function $F$ and saving data needed for correct restoring. In this section, we describe protocols for exactly-once enforcement on top of a lightweight deterministic model for distributed stream processing called {\em drifting state}~\cite{we2018adbis}. This model is implemented in~\FlameStream\ processing system~\cite{we2018beyondmr}.

In~\FlameStream\, computations are deterministic due to a (speculative) maintenance of a pre-defined total order on elements before each order-sensitive operation. The order can be defined using various methods~\cite{we2018seim}, and we assume that $\forall x_1,x_2\in \Gamma, \exists t(x): x_1 < x_2 \Longleftrightarrow t(x_1) < t(x_2)$. In order to achieve exactly-once on top of a deterministic stream processing system we propose the following way:
\begin{itemize}
    \item Periodically save (take a snapshot of) operation states
    \item Consistently restore these states and replay missed input elements in case of failure
    \item Ensure that output elements which have been already released will not be duplicated after recovery and input reprocessing
\end{itemize}

We do not enforce a strict architecture of the system, but assume that there are several functional agents:
\begin{itemize}
    \item Coordinator manages snapshotting and recovery.
    \item Cluster state manager is used to keep service information for recovery.
    \item Persistent storage is needed for reliable storing of state snapshots. In case of failures, the state snapshot is recovered from persistent storage.
    \item Data producer that can replay some set of previous input elements with the same $t(a)$.
    \item Data consumer that is responsible for output elements receiving. The exact requirements for data consumer are detailed further in this section.
    \item Node is an executor of user-defined operations. The computation can be balanced through the distribution of data elements among nodes. Each Node has a {\em barrier} that delivers output elements to the end-user.
\end{itemize}

The implementation of this architecture in~\FlameStream\ processing engine is demonstrated in Figure~\ref{arch}. Apache ZooKeeper is used for cluster state management. As persistent storage, one can use a distributed file system or database (e.g., HDFS, S3, MongoDB, GFS, HBase, etc.). Data producers may vary as well: the role of $t(a)$ can be played by any monotonic sequence, e.g., offsets in Kafka.

\subsection{Snapshotting}

\subsubsection{Operation states}

In order to periodically save operation states, there is a need to determine which input elements affect them. 
 We have to trace reversed dependencies: for each state element $s$ we desire to determine such input elements $\{a_i\}$ that $\forall i, s \in Cl_D^{-1}(a_i)$. Otherwise, it may be unclear, which input elements must be reprocessed during recovery. In Apache Flink, it is implemented through specific streaming elements which play the role of notifications that a determined set of input elements have been processed. In~\FlameStream\ this mechanism is implemented using a modification of {\em Acker} agent from Apache Storm~\cite{apache:storm}. 

Assuming that the mechanism for detection of which input elements influence states is implemented, the protocol of state snapshotting for {\em Coordinator} and {\em nodes} is the following:

\begin{itemize}
    \item Coordinator decides that snapshot should be taken and enables a mechanism to detect which input elements have been processed since the previous snapshot.
    \item On the notification, nodes asynchronously make operation states recoverable and send to Coordinator the acceptance message.
    \item When the Coordinator receives all acceptance messages, it saves the information about which input elements belong to this snapshot and other service information about the snapshot. In general, it is sufficient to save only $t(a)$ of the last input element that affects the snapshot.
\end{itemize}

\begin{figure}[htbp]
  \centering
  \includegraphics[width=0.42\textwidth]{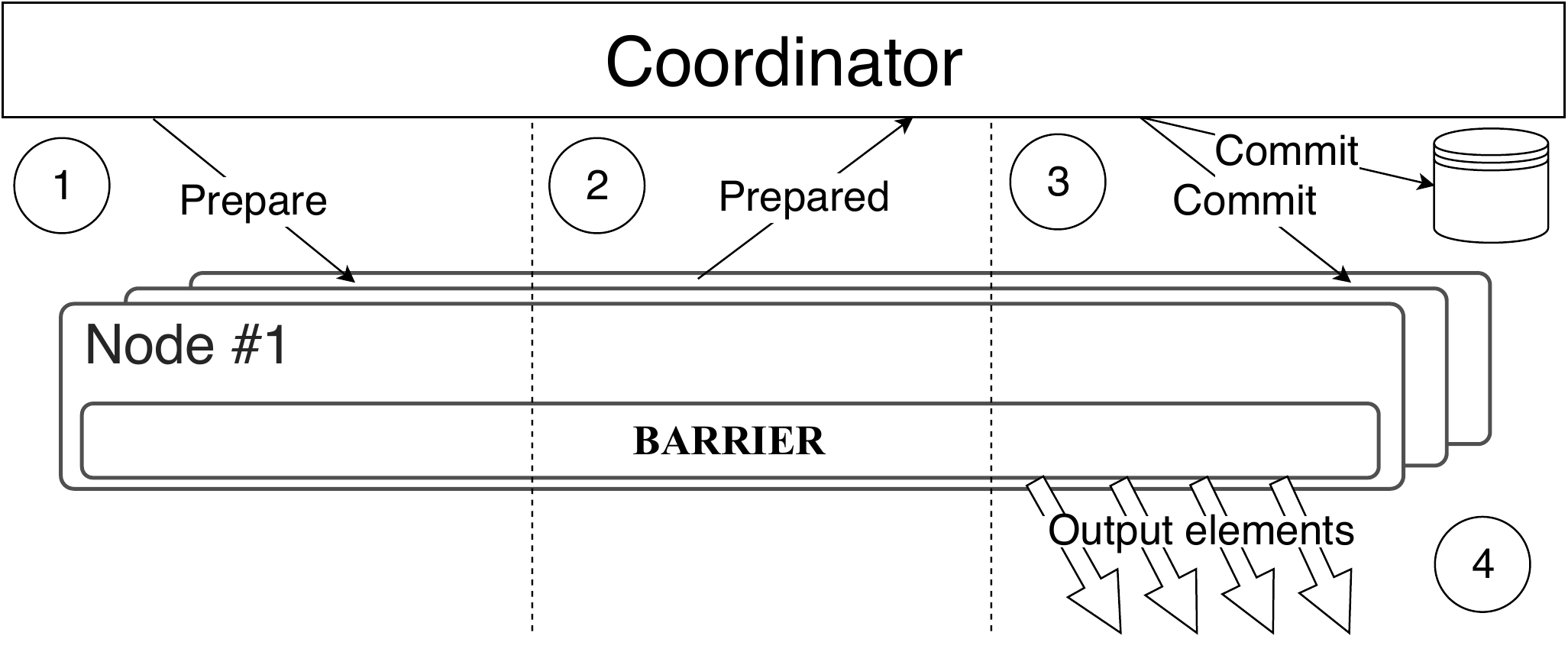}
  \caption{State snapshotting protocol in Apache Flink}
  \label{protocol_flink}
\end{figure}

\begin{figure}[htbp]
  \centering
  \includegraphics[width=0.42\textwidth]{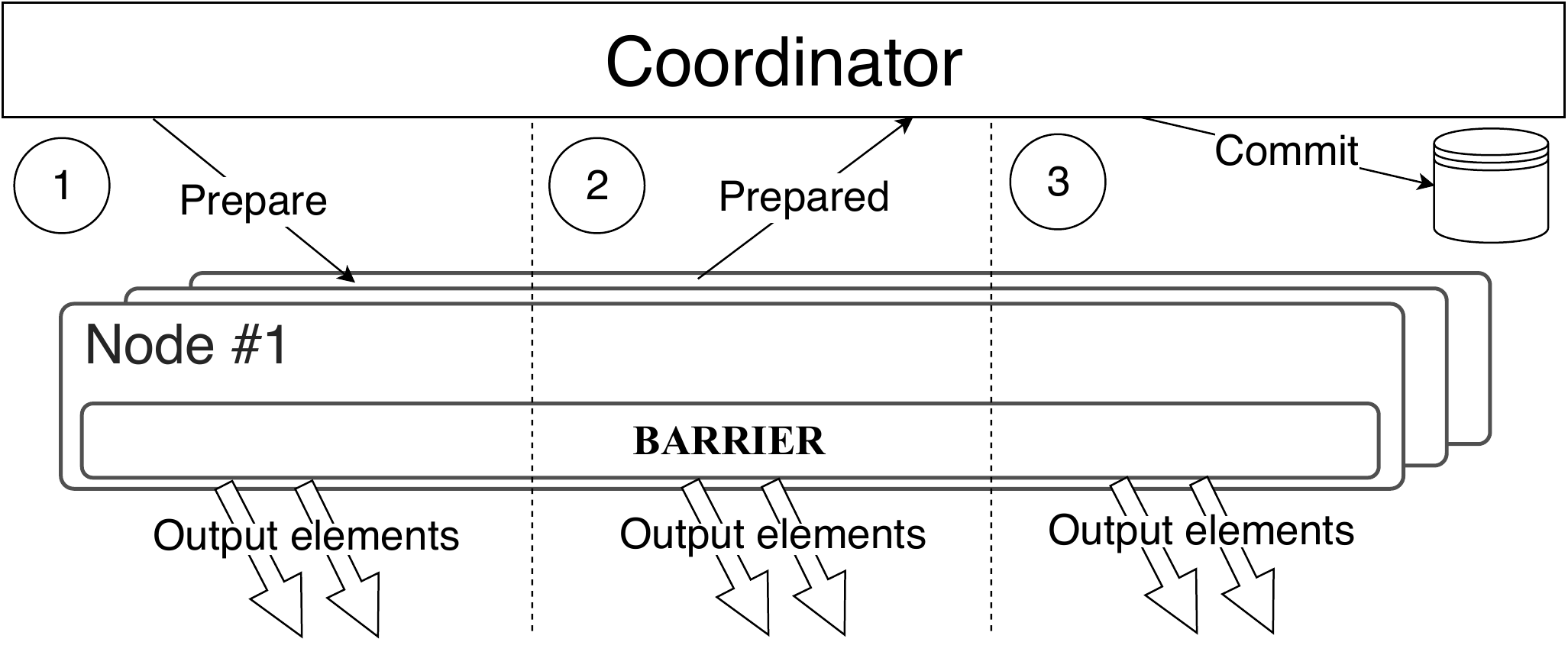}
  \caption{State snapshotting protocol in a deterministic system}
  \label{protocol_fs}
\end{figure}

It is worth to note that the proposed protocol is similar to the transactional (variation of 2PC) state snapshotting protocol used in Flink~\cite{Carbone:2017:SMA:3137765.3137777}. The critical difference is that in our method, output releasing agents (barriers) do not take part in a distributed transaction, because in a deterministic system there is no need to wait until the snapshot is taken in order to release output elements consistently. The scheme of the protocol used in Apache Flink is shown in Figure~\ref{protocol_flink}. As it is demonstrated, output elements delivery (stage 4) is allowed only after commit (stage 3). Modified state snapshotting protocol for a deterministic system is demonstrated in Figure~\ref{protocol_fs}. One can note that elements delivery is independent of the snapshotting protocol. This difference determines the significant latency decrease that is demonstrated further in experiments.

\subsubsection{Last output element}

In order to preserve the deterministic outcome, the Barrier must release output items with monotonically increasing $t(x)$. Hence, the Barrier can filter out any items with $t(x)$ less than or equal to the $t(x)$ of the last released item $t_{last}$ in order to preserve exactly-once after recovery. To implement this mechanism, there is a need to deliver output items and update $t_{last}$ atomically. To solve this problem, we require the following output protocol between a {\em barrier} and a {\em data consumer}: 

\begin{itemize}
    \item On each output delivery, Barrier sends a bundle to the data consumer. This bundle contains an output item and $t_{last}$. The Consumer must acknowledge that it received the bundle.
    \item Barrier does not send new output bundle until the previous one is not acknowledged.
    \item Consumer must return last received bundle on Barrier's request.
\end{itemize}

This protocol guarantees that $t_{last}$ and released items are always consistent with each other. It implies that the Barrier can request the last released bundle and fetch $t_{last}$ after recovery to avoid duplicates, which can be generated during input elements reprocessing.

Thus, on the one hand, we delegate the part of the basic functionality to data consumers. On the other hand, the requirement for a data consumer is not so strong and can be naturally satisfied by real-world consumers (HDFS, Kafka, databases, etc.). 

\subsection{Recovery}

Typically, distributed systems take into consideration the following types of failures:
\begin{itemize}
    \item Packet loss
    \item Node failure
    \item Network partitioning
\end{itemize}

Network partitioning is the particular case of failure because in this case, computations cannot be restarted. We believe that in this case, stream processing does not make sense. To the best of our knowledge, there are no open-source stream processing systems that tolerate network partitioning.

Failure detection can be implemented in different ways~\cite{hayashibara2002failure}, which are out of the scope of this paper. In~\FlameStream\ it is implemented using {\em Acker}~\cite{we2018adbis}. In the case of packet loss or node failure, a failure detector mechanism can enforce the Coordinator to begin computations restart from the last successful snapshot. Restart protocol includes the following steps:

\begin{itemize}
    \item Coordinator broadcasts a notification to begin the recovery process.
    \item Operations receive these notifications and fetch their states from state storage. After that, they send an acknowledgment that they are ready for Processing to the Coordinator.
    \item Barriers request the last released bundle from data consumers and send acknowledgments that they are ready for Processing to the Coordinator.
    \item When the Coordinator receives all acknowledgments from groupings and barriers, it requests data producer to replay starting from the $t(a)$ of the last snapshot.
\end{itemize}

The proposed protocol guarantees the following properties that allow preserving~ exactly-once:

\begin{itemize}
    \item Processing does not restart until all operations obtain consistent states. The consistency of these states is guaranteed by the state snapshotting protocol. Therefore, the elements are not lost.
    \item Duplicates are not produced because, at the moment when Processing is restarted, it is ensured that Barrier has obtained the last released $t(x)$ and can filter out extra items.
\end{itemize}

\section {Experiments}

\label {fs-experiments-seciton}

\subsection{Setup}
A series of experiments were performed in order to analyze the overall performance of the proposed approach. For experiments, we use an open-source implementation of the drifting state model called~\FlameStream. \FlameStream\ is a distributed stream processing engine implemented in Java using Akka framework. \FlameStream\ can be deployed on a hardware cluster of computational units that we call nodes. We assume that each node is connected through a network with all other nodes.

\begin{figure}[htbp]
  \centering
  \includegraphics[width=0.50\textwidth]{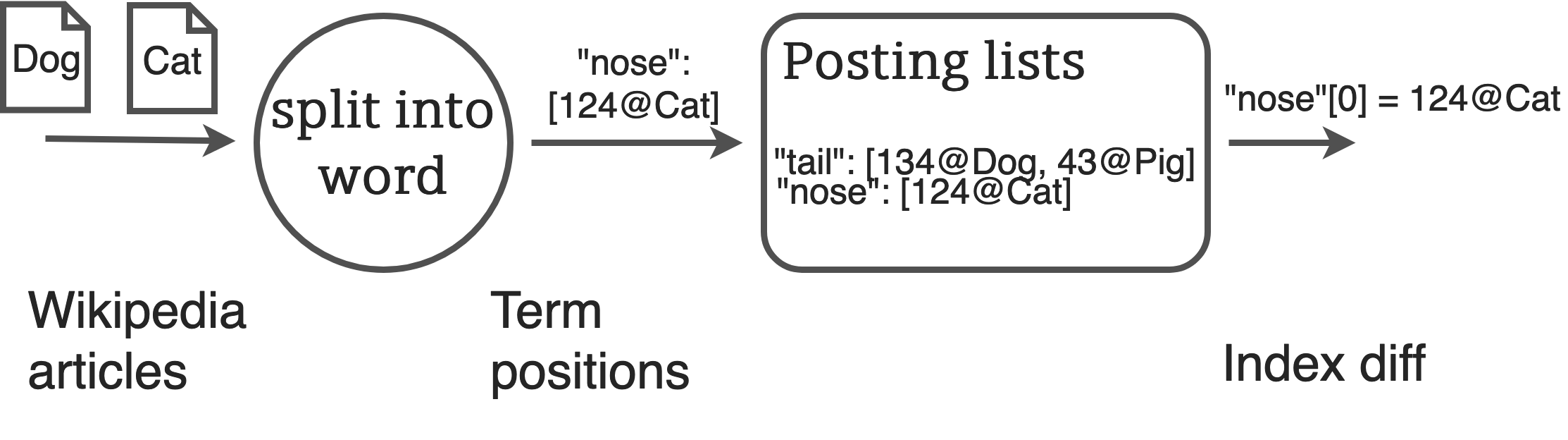}
  \caption{The inverted index pipeline}
  \label {index}
\end{figure}

We used a problem of incremental inverted index maintenance over a stream of text items. 
Building inverted index is implemented as a MapReduce transformation in a streaming manner. The scheme is shown in Figure~\ref{index}: 

\begin{itemize}
    \item Map phase includes conversion of input documents into the key-value pairs {\it (word; word positions within the page)}
    \item Reduce phase consists of combining word positions for the corresponding word into the single index structure 
\end{itemize}

The Reduce phase produces the change records of the inverted index structure, to make this algorithm suitable for stream processing systems. It implies that each input page triggers the output of the corresponding change records of the full index. In \FlameStream\ this algorithm is implemented as the typical conversion of MapReduce transformation, which is shown in~\cite{we2018seim}.

We choose the problem  of  an inverted index maintenance  because it satisfies the following properties:

\begin{itemize}
    \item Operation that generates change records is non-commutative
    \item The computational workflow    contains network shuffle that can violate the ordering constraints
    \item Consistency guarantees are strongly required because the inconsistent index does not make sense for many applications
    \item The workload is unbalanced due to Zipf's law
\end{itemize}

Notably, building an inverted index in a streaming manner can be the halfway task between the generation of documents and consuming index updates by search infrastructure. In the real world, this scenario can be found in freshness-aware systems, e.g., news processing engines.

By the notion of {\it latency} we assume the time between two events: 

\begin{enumerate}
    \item Input page is taken into the stream
    \item All the change records for the page leave the stream
\end{enumerate}

Our experiments were performed on the cluster of 10 Amazon EC2 micro instances with 1GB RAM and 1 core CPU. We used Wikipedia articles as a dataset. Documents per second input rate is 50 because higher rates lead to enabling of backpressure mechanisms in an industrial system that we compare with. RocksDB~\cite{rocksdb} is used as a storage for the state. The role of data producer and data consumer is played by a custom server application that sends and receives data through socket and measures the latency.

\subsection{Overhead and recovery}
The performance of the proposed deterministic model within the same stream processing task is deeply analyzed in~\cite{we2018seim}. In this paper, we aim to evaluate the overhead on providing consistency guarantees and the time needed for the full recovery.

Figures~\ref{comparison50}, ~\ref{comparison500}, and ~\ref{comparison1000} show the latencies of \FlameStream\ within distinct times between checkpoints. As expected, the overhead on exactly-once enforcement is low (less than 10 ms), and it does not depend on the time between checkpoints. Slight overhead can be explained by the fact that asynchronous state snapshotting is executed on single-core nodes. The time between checkpoints does not influence latency because output elements delivery and state snapshotting mechanisms are independent in our model.

\begin{figure}[htbp]
  \centering
  \includegraphics[width=0.48\textwidth]{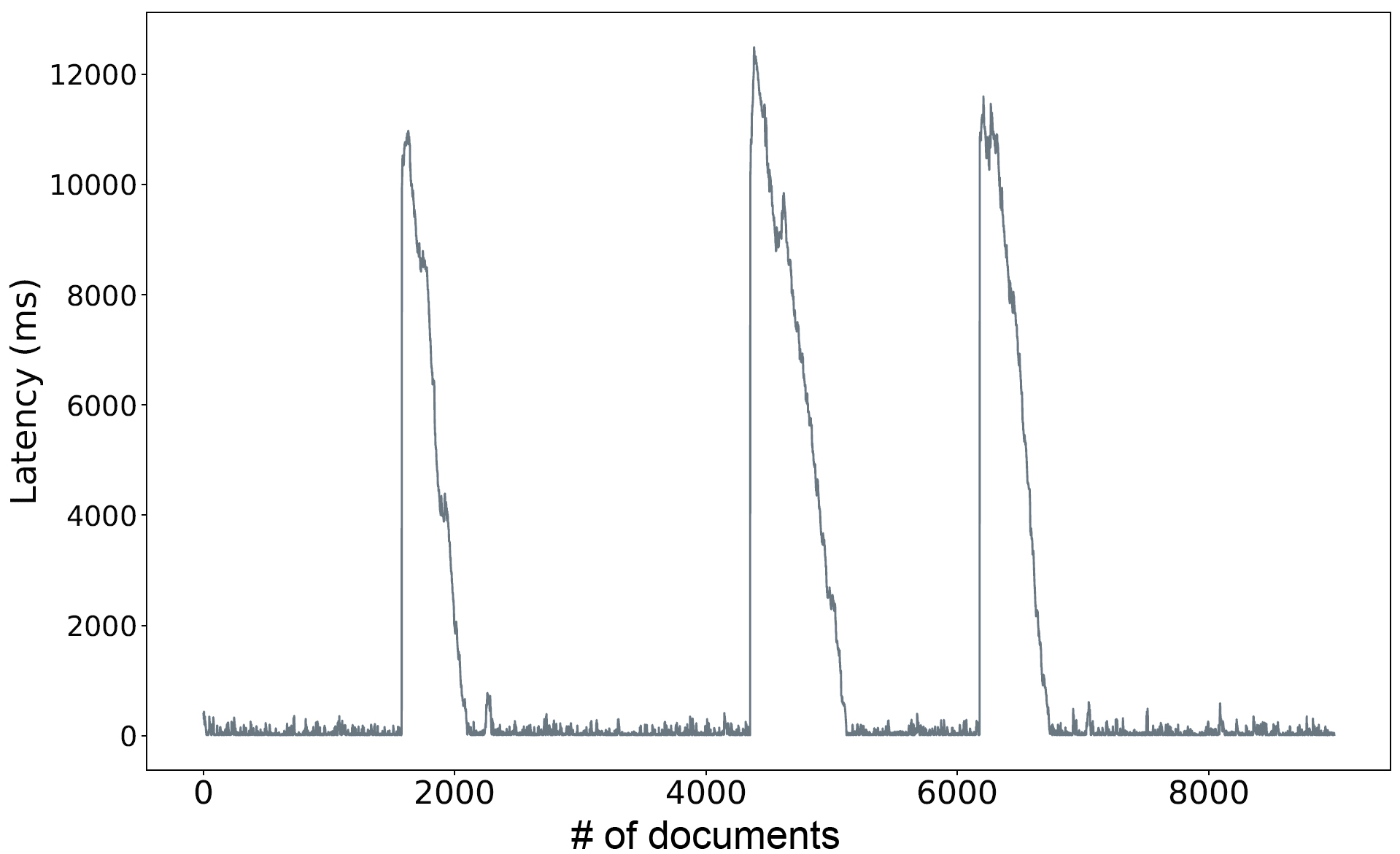}
  \caption{The latencies of \FlameStream\ during three artificially reproduced failures and recoveries with 1000 ms between checkpoints}
  \label {recovery}
\end{figure}

The system behavior in case of failures and recoveries with 1000 ms between checkpoints is demonstrated in Figure~\ref{recovery}. It is shown that the system can perform recovery processes in an adequate time. Existing latency spikes are caused by the replay process, JVM restart, etc.

\subsection{Comparison with an industrial system}
One of the most important goals of the experiments is the performance comparison with an industrial solution regarding latency. Apache Flink has been chosen for evaluation because it is a state-of-the-art stream processing system that provides similar functionality and achieves low latency in     q             the real-world scenarios~\cite{S7530084}. 

For Apache Flink, the algorithm for building the inverted index is relies on the usage of {\it FlatMapFunction} for map step and stateful {\it RichMapFunction} for reduce step and for producing the change records. The network buffer timeout is set to 0 to minimize latency. Custom {\it TwoPhaseCommitSinkFunction} that buffers output items in memory until a transaction is committed is used for experiments that require exactly-once semantics. 

{\it FsStateBackend} with the local file system is used for storing the state, because {\it RocksDBStateBackend} requires saving state to RocksDB on each update that leads to additional overhead. {\it FsStateBackend} stores state on the disk only on checkpoints and do not provide an additional overhead against RocksDB storage in \FlameStream, so it is fairer to use it rather than {\it RocksDBStateBackend} for comparison purposes.

In this paper, we compare $50^{th}$, $75^{th}$, $95^{th}$, and $99^{th}$ percentile of distributions, which clearly represent the performance from the perspective of the users' experience.

\begin{figure}[htbp]
  \centering
  \includegraphics[width=.48\textwidth]{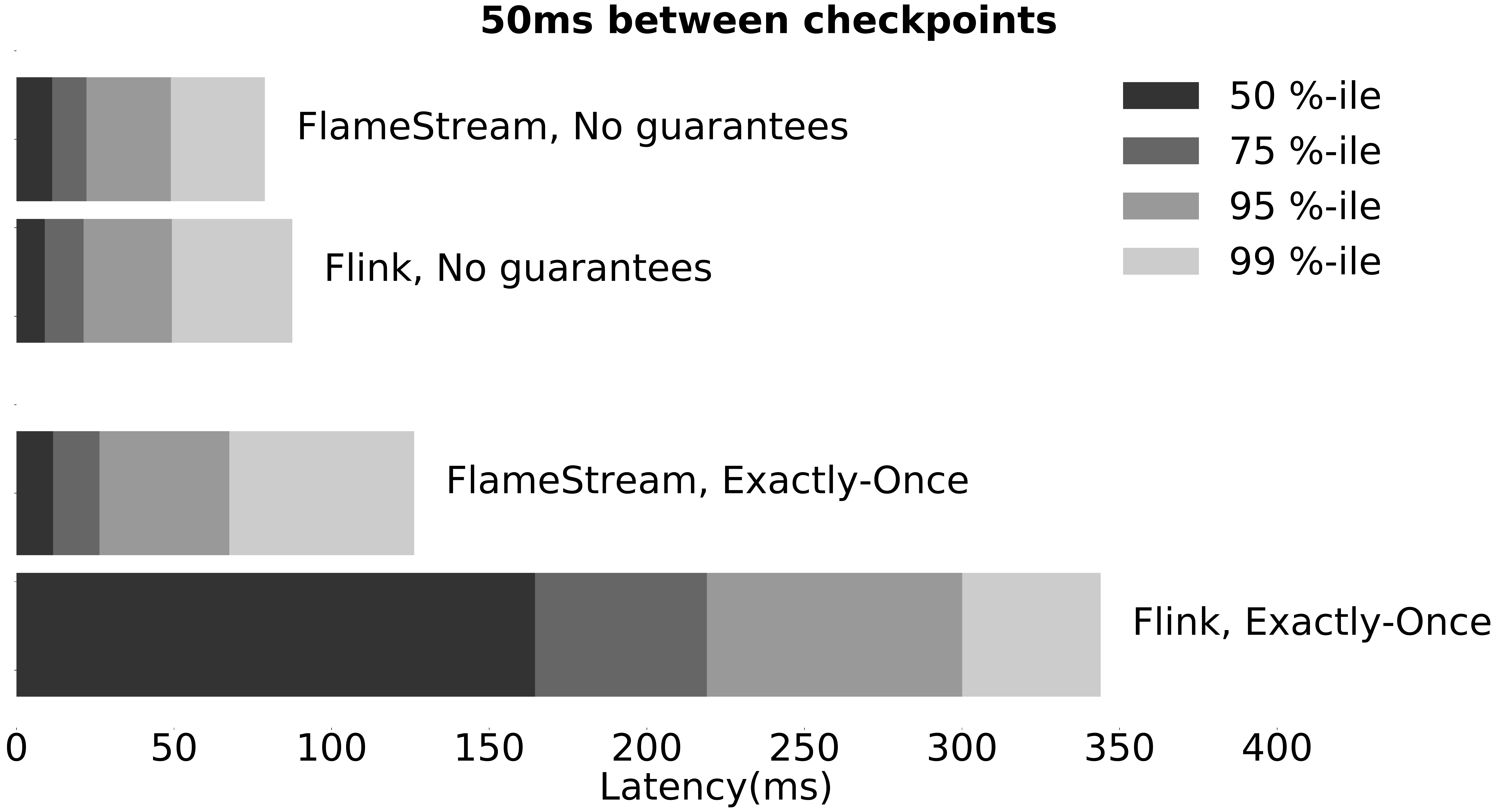}
  \caption{\FlameStream\ and Flink latencies with 50ms between checkpoints}
  \label{comparison50}
\end{figure}

\begin{figure}[htbp]
  \centering
  \includegraphics[width=.48\textwidth]{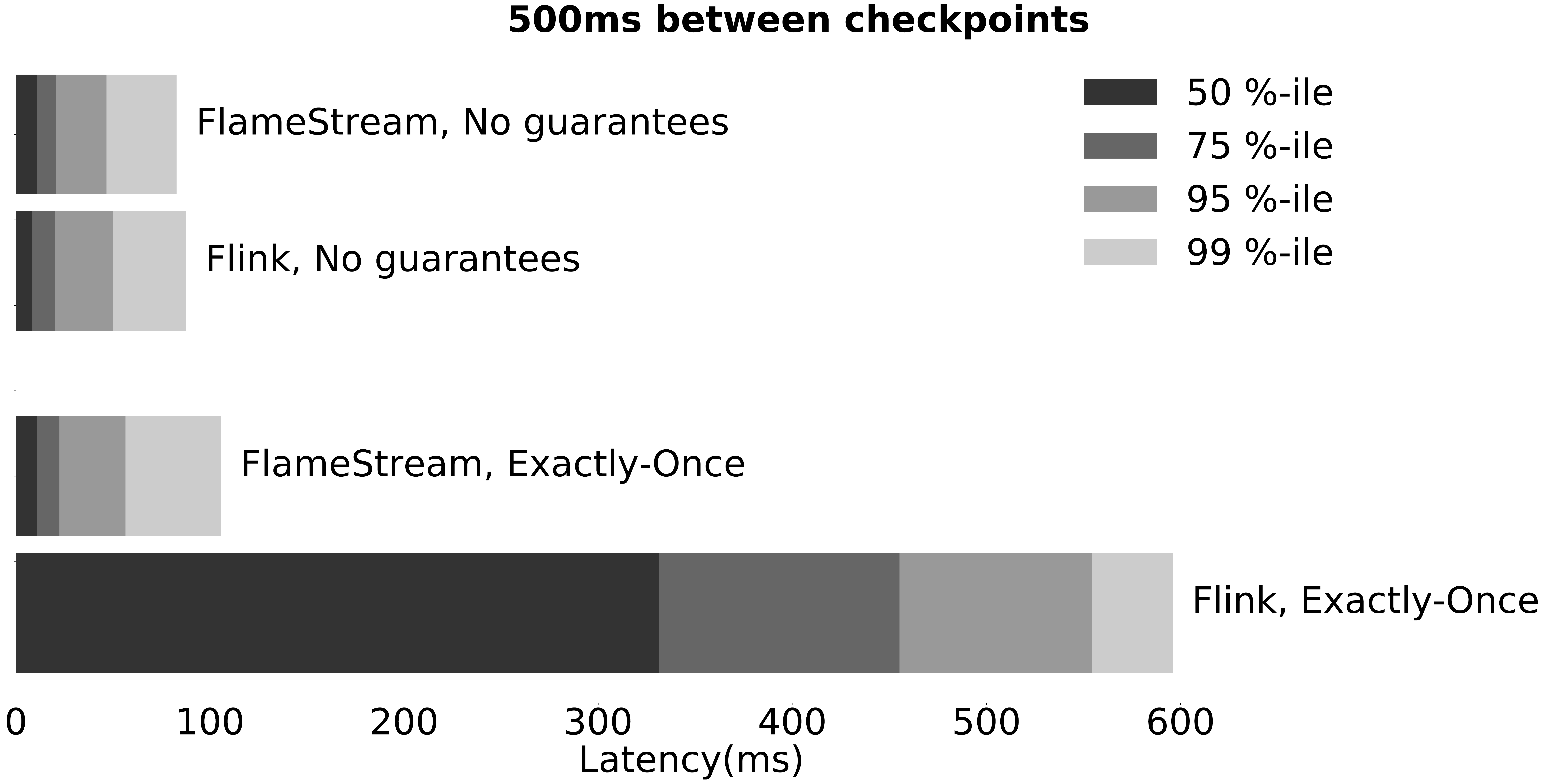}
  \caption{\FlameStream\ and Flink latencies with 500ms between checkpoints}
  \label{comparison500}
\end{figure}

\begin{figure}[htbp]
  \centering
  \includegraphics[width=.48\textwidth]{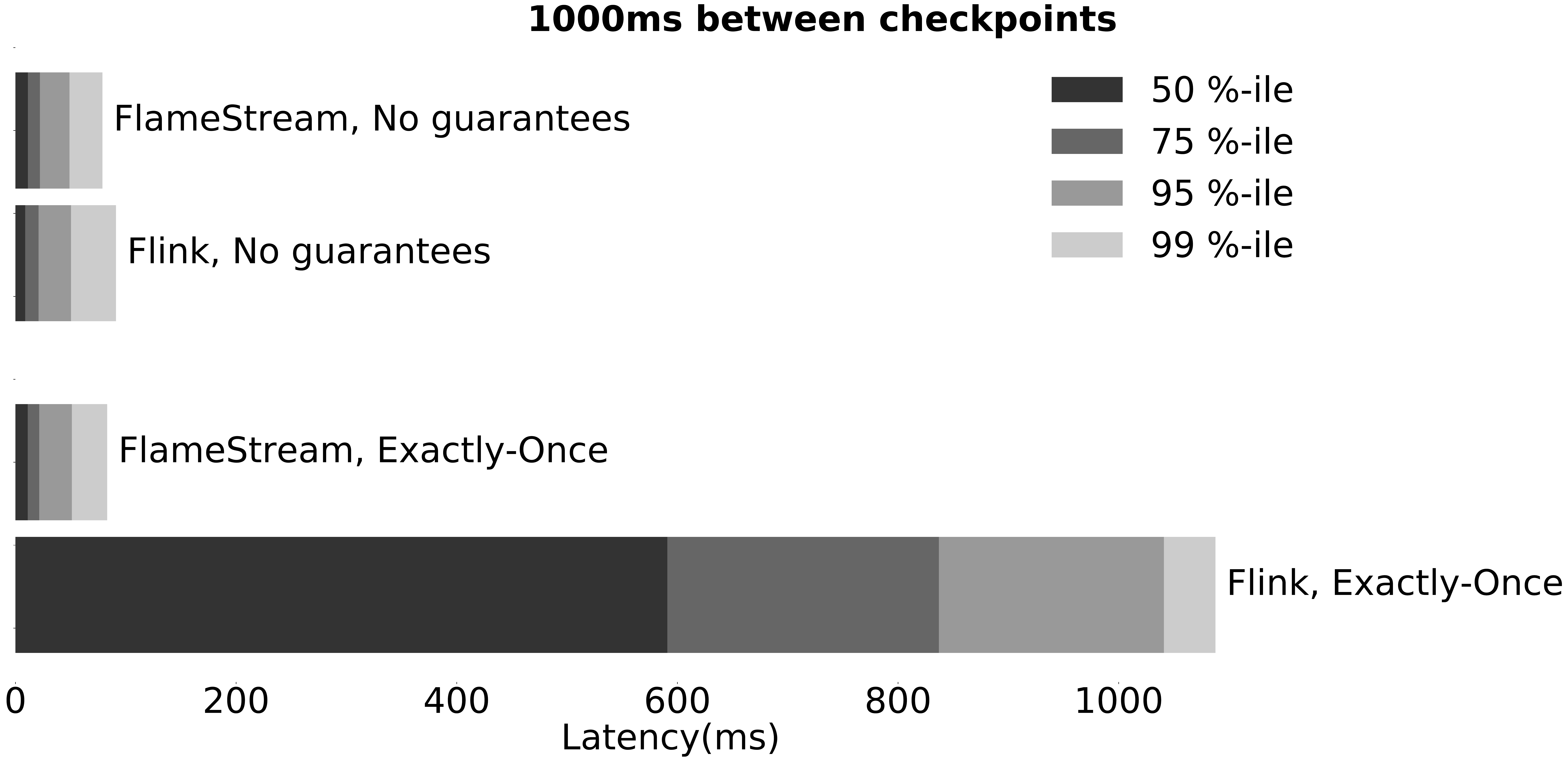}
  \caption{\FlameStream\ and Flink latencies with 1000ms between checkpoints}
  \label{comparison1000}
\end{figure}

Figures~\ref{comparison50}, ~\ref{comparison500}, and ~\ref{comparison1000} demonstrate      the comparison of latencies between \FlameStream\ and Flink within distinct times between checkpoints, and different guarantees on data. Without guarantees, the latencies of \FlameStream\ and Flink do not significantly differ. However, for exactly-once, Flink's latency is dramatically higher, and it directly depends on the time between checkpoints. Nevertheless, such behavior is expected, because Flink needs to take state snapshot and release output items within a single transaction to ensure that all states of non-commutative operations are persistently saved. There are no hints implemented in Flink which could mark an operation as commutative, hence it waits until states of all operations are stored before output delivery. 
On the other hand, the property of determinism allows \FlameStream\ to avoid synchronization state snapshotting and output delivery. This fact makes it possible to achieve exactly-once with low overhead.

\section {Related work}

\label {fs-related-seciton}

Techniques for providing exactly-once~\cite{Carbone:2017:SMA:3137765.3137777, Akidau:2013:MFS:2536222.2536229, Zaharia:2012:DSE:2342763.2342773} are discussed   in details in sections~\ref{fs-intro-seciton} and~\ref{fs-eo-impl}. Many prior works in the field of stream processing do not consider exactly-once maintenance. 
For instance, Aurora~\cite{Abadi:2003:ANM:950481.950485} and Borealis~\cite{abadi2005design} do not provide any guarantees on data at all. Some other systems provide only partial consistency. Apache Storm~\cite{apache:storm} supports message tracking mechanism that prevents the loss of data. 
However, exactly-once semantics is not provided, because duplicates are still possible. Twitter Heron, that was presented as the next generation of Apache Storm~\cite{Kulkarni:2015:THS:2723372.2742788}, does not provide for exactly-once as well. 
Samza~\cite{Noghabi:2017:SSS:3137765.3137770} also implements fairly similar to Storm model and has the same consistency guarantees.

Prior works on stream processing formalization concentrate on operations specification rather than delivery guarantees. Logical foundation for specifying streaming computations is discussed in~\cite{alur2018interfaces}. Declarative algebraic notations for the streaming queries are introduced in~\cite{halle2014formalization}. Another direction in streaming formalization is designing frameworks to define operations semantics~\cite{beck2018lars}.

\section {Conclusion}

\label {fs-conclusion-seciton}

We introduced a formal, conceptual framework for modeling consistency properties for any stream processing system. We demonstrated how the behavior of state-of-the-art research and industrial systems could be described in terms of the proposed framework. It was shown that the property of determinism is tightly connected with the concept of of exactly-once. We proved that non-deterministic systems must persistently save a state of non-commutative operations before output delivery in order to achieve exactly-once. Most of the state-of-the-art stream processing systems use one of the following approaches to overcome this problem: 

\begin{itemize}
    \item Inherit exactly-once from batch processing using small-sized batches (micro-batching)
    \item Apply distributed transaction control protocols which guarantee that states are saved before delivery of elements affected by these states
    \item Write results of an operation to external storage on each input element
\end{itemize}

All these methods experience difficulties with working under low-latency requirements (less than a second). In the first case, latency cannot be lower than the batching period, in the second case, the distributed two-phase commit may result in a significant increase of latency, while in the third case latency is bounded below by the duration of external writes.

Using our formal inference, we designed mechanisms for achieving exactly-once on top of {\em drifting state} technique introduced in our previous work~\cite{we2018adbis}. Drifting state provides inexpensive determinism due to optimistic nature and low overhead on a single buffer per any stateful data flow. Because of the determinism, the protocols provide the following features:

\begin{itemize}
    \item Elements are processed in a pure streaming manner without input buffering
    \item The processes of business-logic computations, state snapshotting and delivery of output items work asynchronously and independently
    \item Exactly-once is preserved
\end{itemize}

We implemented the prototype of the proposed technique to examine its performance. Our experiments demonstrated that the introduced protocols for fault tolerance are scalable and provide remarkably low overhead within different computational layouts. Furthermore, the comparison with the industrial stream processing solution indicated that our prototype could provide significantly lower latency under exactly-once requirement.

\balance

\bibliographystyle{IEEEtran}
\bibliography{flame-stream}

\end {document}